

\input harvmac

\Title{BROWN-HET-932, hep-th/9401134}{High Energy Scattering in 2+1 QCD }
\centerline{Miao Li\foot{E-mail: li@het.brown.edu}
and Chung-I Tan\foot{E-mail: tan@het.brown.edu}}
\bigskip
\centerline{\it Department of Physics}
\centerline{\it Brown University}
\centerline{\it Providence, RI 02912}
\bigskip
High energy scattering in 2+1 QCD is studied using the recent approach of
Verlinde and Verlinde. We calculate the color singlet
part of the quark-quark scattering exactly within this approach, and discuss
some physical implication of this result. We also demonstrate, by two
independent methods, that reggeization fails for the color octet channel.
We briefly comment on the problem in 3+1 QCD.

\Date{1/94}

\newsec{Introduction}

Quantum chromodynamics (QCD) with massless quarks is classically scale
invariant; it contains no explicit  small parameters. Quantum mechanically, the
character of  QCD changes depending on  the nature of available  probe. At
short distances, due to asymptotic freedom, the basic degrees of freedom are
weakly coupled quarks and gluons.  Collisions where all
 invariants $\{p_i{\cdot}p_j\}$ are large probe short-distance
physics and can be treated perturbatively. Over the past two
decades, perturbative tests of QCD have been met with continued success,
leaving little doubt on it being the correct theory of strong interactions. As
one moves to larger distance scales, the coupling strength increases and one
enters the nonperturbative quark-gluon confinement regime. Short of resulting
to lattice numerical studies, the most promising tool for a non-perturbative
treatment of QCD which builds in confinement naturally remains the topological
treatment based on
$1/N_c$ expansion.  Indeed,  many qualitative features of  low energy
hadronic  phenomenology can best be understood in such a setting.
Unfortunately, quantitative calculational scheme is still lacking, and it is
unlikely one could be developed in the near future.

Formally,  a   high energy hadronic collision in the near-forward
limit corresponds to the mixing of  a ``short-distance"
phenomenon in the longitudinal coordinates with  a ``long-distance" phenomenon
in the transverse coordinates. By treating the longitudinal and transverse
degrees of freedom separately, one could hope that  a ``dimensional
reduction" scheme can be formulated, reducing QCD at high energies to an
effective two-dimensional field theory. An interesting attempt in this
direction has been made recently by Verlinde and Verlinde \ref\vv{H. Verlinde
and E. Verlinde, preprint PUPT-1319, revised version.}.  The effective
theory involves several fields and two coupling constants, the original gauge
coupling
$g$ and an effective coupling $e^2 \sim g^2\log s$.

In this paper, we
would like to explore the viability of this approach further by study a
simpler situation, high energy scattering of QCD in 2+1 dimensions, where the
resulting effective theory is one-dimensional.

In Ref.\vv, one first accepts the conventional wisdom that the high energy
behavior of  hadron-hadron near-forward amplitudes in QCD can be extracted by
studying  the corresponding quark-quark scattering amplitudes. One further
assumes that the latter can  be expressed in terms of correlators of
certain Wilson lines \ref\nach {O. Nachtmann, Ann. Phys. 209 (1991) 436.}.
Denote the center of mass squared by $s$ and the momentum transfer squared by
$t$; one is interested in the region where $s>>|t|$. One normally also assumes
that
$|t|$ is still greater than $\Lambda_{QCD}$. This later condition
hopefully ensures that the perturbation makes sense. As we shall see,
Verlindes' formulation presumes no perturbative expansion, and indeed
explicit nonperturbative calculation can actually be done in this approach
when one considers 2+1 QCD. For related approaches, see \ref\irina{
I.Ya. Aref\'eva, preprint SMI-5-93, hep-th/9306014; SMI-15-93,
hep-th/9311115; S.-J. Rey, talk given at 5th Blois Workshop on Elastic and
Diffractive Scattering, hep-ph/9308332.}.

Let $\alpha$ be the longitudinal
index and $i$ the transverse index. Starting with the standard Yang-Mills
action, let us  scale the longitudinal coordinates by a factor
$\lambda=1/\sqrt{s}$, which corresponds to a scaling of
longitudinal momenta by  $\sqrt{s}$. It follows that  components
$F_{\alpha\beta}$ of the field strength are scaled by a factor
$\lambda^{-2}$, $F_{\alpha i}$  by a factor $\lambda^{-1}$, with
$F_{ij}$ left intact, {\it i.e.}, the Yang-Mills action now reads as
\eqn\res{S={1\over 4g^2}\int d^4x\hbox{tr}(\lambda^{-2}F_{\alpha\beta}^2+
2F^2_{\alpha i}+\lambda^2F_{ij}^2).}
Verlinde and Verlinde  first assume that, with  $\lambda$
small at high energies, one can neglect the third term in the above
action. We note here that in working with 2+1 dimensions,
this term is actually absent, because there is only one transverse dimension.
It thus removes one possible sources of uncertainties concerning this novel
approach \ref\polish{L.N. Lipatov, Nucl. Phys. B365 (1991) 614; R. Kirschner,
hep-th/9311159.}.

The next crucial assumption made in \vv\ is that the fluctuation of the
longitudinal components is suppressed due to  the first term in the above
equation.  $A_\alpha$ is therefore flat in the longitudinal directions and
one  can replace it by $\partial_\alpha G G^{-1}$ in the
second term. Finally, one integrates over longitudinal coordinates, leading to
an effective two dimensional action, involving seven matrix fields.
Unfortunately, this effective theory is still too complicated to be
amenable to explicit calculation. As we shall see, by studying 2+1 QCD, our
effective theory is one-dimensional, involving only three independent fields.
This allows us to study the model both perturbatively as well as exactly in
$e^2(s)$.

In order to provide some relevant background for proper appreciation of the
possible significance of the  approach of Ref.\vv, we recall that one of
the most striking aspects of high-energy hadron-hadron scattering is the
continued increase of the total cross section
$\sigma_T$ with the energy. Traditional approach to high energy near-forward
hadronic collisions invariably involves the notion of Pomeron. A rising
cross section first requires a Pomeron with a zero momentum transfer
intercept greater than one, which, if uncorrected, would lead to the
violation of Froissart bound at asymptotic energies. Either through an
eikonal or other more elaborate schemes, screening corrections hopefully
would then lead to an expanding disk picture.

There are currently two
seemingly conflicting interpretations of Pomeron in QCD: One  based on
perturbative leading log approximation (LLA)
\ref\Lipatov{L.N. Lipatov, review in {\it Perturbative QCD}, ed. A.H.
Mueller (World Scientific, Singapore, 1989), and references therein.}
\ref\cw{H. Cheng and T.T. Wu, Expanding Protons: Scattering at High Energies,
The MIT Press (1987).} \ref\lip{E. Kuraev, L.N. Lipatov and V. Fadin, Sov.
Phys. JETP 44 (1976) 443; 45 (1977) 199; Ya. Balitski and L.N. Lipatov,
Sov. Nucl. Phys. 28 (1978) 822.}
\ref\Leningrad{L.V. Gribov, E.M. Levin, and M.G. Ryskin, Phys. Rep.
  100C (1983) 1; E.M. Levin and M.G. Ryskin, Phys. Rep.  189C (1990)
267.}
and another  based on nonperturbative (large-$N$ and/or phenomenological)
consideration \ref\DPM{A. Capella, U. Sukhatme,
C-I Tan and Tran T. V., Phys. Lett.  B81 (1979) 68;
see also: {\it Dual Parton Model}, Phys. Rep. 236 (1994) 225.}
\ref\Landshoff{P. V. Landshoff, Proc. of 3rd
Int. Conference on Elastic and Diffractive Scattering, Nucl. Phys.  B12
(1990) 397.}. In a perturbative treatment, a Pomeron loosely corresponds to
the color-singlet bound state of  two (reggeized) gluons. In a
nonperturbative treatment, the Pomeron is thought  to correspond to
nonperturbative gluon exchanges having the topology of a closed string.

One potential advantage of the Verlindes' approach, as stated earlier,
is the fact that  it offers the
possibility of providing either an alternative  perturbative LLA treatment or
a starting point for nonperturbative studies. An attempt in unifying
both perturbative and nonperturbative aspects of high energy hadron collisions
has been made in \ref\letan{E.M. Levin and C-I Tan, Heterotic Pomeron: High
Energy Hadronic Collisions in QCD, in preparation; a brief outline can
be found in Proc. of XXII Int. Symposium on Multiparticle
Dynamics, World Scientific (1992).}.
In generating a perturbative Pomeron \Lipatov\ \Leningrad, it is important to
identify on the one hand the ``Lipatov
vertex" and on the other hand the reggeization of gluons. The former has been
demonstrated in Ref.\vv, but the possibility of gluon reggeization remains
unclear. In section 3 we demonstrate that, under Verlindes' approach,
reggeization does not occur for 2+1 QCD at high energies. This may pose
a question as whether one need modify the ansatz for how $\log s$ enters
the effective action.

Although the above result might cast doubts on the reliability of the
Verlindes' scheme, it is nevertheless interesting to push the program further
for extracting other nonperturbative consequences. We begin first by
introducing a new procedure in section 4 which allows to perform
nonperturbative analysis. In section 5 we discuss exact result for the
quark-quark high energy scattering for 2+1 QCD under Verlindes'  treatment. In
section 6 we discuss hadron-hadron scattering in 2+1 QCD and briefly comment
on the case of 3+1 QCD.

\newsec{Verlindes' approach to high energy scattering in QCD}

In this section we first briefly review Verlindes' approach \vv\ to high
energy quark-quark scattering at fixed momentum transfer. We then simplify
their effective action to set the framework for calculation of scattering
amplitude.

Consider quark-quark scattering in QCD with center of mass energy squared
$s$. $s$ is very large, and the momentum transfer $t$ is much smaller than
$s$, but still greater than $\Lambda_{QCD}$. This later condition ensures
that the perturbation makes sense. Indeed, as we shall see, Verlindes'
formulation presumes no perturbative expansion, and indeed nonperturbative
calculation can be done in this formulation when one considers 2+1 QCD.

Under the same assumption where one drops lower order terms under the
longitudinal scale transformation, the quark-quark scattering amplitude can be
expressed in terms of correlation of two Wilson lines
\eqn\ampl{A(s,t=-q^2)=-{is\over 2m_q^2}\int d^2ze^{-iqz} \langle
(V_+(0)-1)(V_-(z)-1)\rangle,}
where $V_\pm =e^{\int^\infty_{-\infty}dx^\pm A_{\pm}},$
 involve longitudinal gauge field components only.
The pre-factor ${is\over 2m_q^2}$ comes from kinematics in longitudinal
dimensions, $m_q$ is the quark mass (we assume the two quarks have the same
mass, although this is not necessary). A detailed justification of the use of
Wilson lines for quark-quark scattering can be found in Ref. \nach.

As stated in the introduction, we accept that it is justifiable at high
energies  to approximate $A_{\alpha}$ by the pure gauge condition
$\partial_\alpha G G^{-1}$ while dropping the first term
$\lambda^{-2}F_{\alpha\beta}^2$ in Eq.\res. One next integrates over
longitudinal coordinates, assuming that the gauge transformed
$A_i$ satisfies classical equation $\partial_+\partial_-(G^{-1}D_iG)=0$.
An effective two dimensional action is thus obtained, involving fields
at four end-points of two quark trajectories. There are totally six fields
in this action. $g_A$, $A=1,2$, are values of $G$ at the two ends of the left
moving quark. $h_A$, $A=1,2$, are values of $G$ at the two ends of the right
moving quark. $a^\pm_i$ are gauge fields associated with the left moving and
the right moving trajectories respectively.

This effective action is singular in that some fields have a singular
propagator. Verlindes then regularize the action by noting the fact that
both quarks are actually not exactly light-like. Their classical
trajectories
depart from their light-cones by an amount proportional to $1/\sqrt{s}$. The
propagator of transverse components $A_i$ then acquires $s$ dependence through
$\hbox{log}s$. We shall not go through their derivation, but just quote
their result. The regularized action is
\eqn\act{S[g_A,h_B, a^{\pm}_i]={1\over g^2}\int d^2zM^{AB}\hbox{tr}
(g_A^{-1}D^+_ig_Ah^{-1}_BD^-_ih_B),}
where the matrix $M$ is
\eqn\prop{\eqalign{&M=\left(\matrix{1+\epsilon &-1+\epsilon\cr -1+\epsilon &1+
\epsilon}\right)\cr
&\epsilon^{-1}=1-{2i\over\pi}\hbox{log}s,}}
and covariant derivatives are given by $D^+_ig_A=(\partial_i+a^+_i)g_A$,
$D^-_ih_A=(\partial_i+a^-_i)h_A$.
Here a comment about the coefficient $1/g^2$ in the action \act\ is in order.
We shall compare result of perturbative calculations based on \act\ to the
conventional perturbation results, so we need be careful in setting the same
convention about the coupling constant. From the rescaled action \res, the
weighting coefficient of the second term is $1/(2g^2)$. A factor $2$ comes from
writing the second term in the light-cone coordinates and integrating out
the longitudinal coordinates.

The  Wilson lines operators
$V_\pm$ introduced earlier for calculating the quark-quark scattering
amplitude can  be written in terms of
$g_A$ and
$h_A$, since
$A_\pm={1\over g}\partial_\pm G G^{-1}$. Let $g=g_2g_1^{-1}$ and
$h=h_2h^{-1}_1$, then $V_+=g$, $V_-=h$. The quark-quark amplitude then becomes
\eqn\ampltwo{A(s,t=-q^2)=-{is\over 2m_q^2}\langle (g-1)(q)(h-1)(-q)\rangle.}
We shall next  simplify the effective action \act\ before discussing
the correlation of these two Wilson line operators.

 So far we have not fixed
any gauge.  Consider the $a^\pm$ independent part of \act\ first:
$$\eqalign{S(g,h)&={1\over g^2}\int d^2z[(1+\epsilon)\hbox{tr}(g^{-1}_2
\partial_i g_2-g^{-1}_1\partial_i g_1)(h^{-1}_2\partial_i h_2-h^{-1}_1
\partial_i h_1)\cr
&+2\epsilon\hbox{tr}[g^{-1}\partial_i g_1(h^{-1}_2\partial_i h_2-h^{-1}_1
\partial_i h_1)+ h^{-1}_1\partial_i h_1(g^{-1}_2\partial_i g_2-g^{-1}_1
\partial_i g_1)]\cr
&+4\epsilon\hbox{tr}(g^{-1}_1\partial_i g_1h^{-1}_1\partial_i
h_1) ].}$$
It is easy to see from the above equation that fluctuations of
$g^{-1}_2\partial_i g_2-g^{-1}_1\partial_i g_1$ and $h^{-1}_2\partial_i h_2-
h^{-1}_1\partial_i h_1$ are controlled by $g^2$, and fluctuations of
$g^{-1}_1\partial_i g_1$ and $h^{-1}_1\partial_i h_1$ are controlled by
$g^2\epsilon^{-1}$, a parameter much greater than $g^2$. Therefore, the
second term
in the above action is negligible compared to the first term and the third
term. Next, observe that
$$4\epsilon \hbox{tr}(g^{-1}_1\partial_ig_1h^{-1}_1\partial_i h_1)=
\epsilon\hbox{tr} \left( (g^{-1}_1\partial_ig_1+h^{-1}\partial_i h_1)^2-
(g^{-1}_1\partial_ig_1-h^{-1}_1\partial_ih_1)^2\right),$$
and $g^{-1}_1\partial_ig_1+h^{-1}_1\partial_ih_1$ essentially decouples
from the action, if we ignore the second term in the action. Now recalling
the definition $g=g_2g^{-1}_1$, $h=h_2h^{-1}_1$ and defining $G=g_1h^{-1}_1$,
the $a^\pm$ independent part of the effective action is simplified to
\eqn\indep{S[g,h,G]={1\over g^2}\int d^2z[\hbox{tr}(g^{-1}\partial_igG
h^{-1}\partial_ihG^{-1})-\epsilon\hbox{tr}(G^{-1}\partial_iG)^2],}
where we have ignored the term proportional to $\epsilon$ in the first
term in \indep.
It is seen that this action is written in terms of $g$, $h$, ``physical''
fields associated with Wilson lines, and $G$ which plays an important role
as a coupling field. In 2+1 QCD, \indep\ is the whole effective action,
if we choose the Landau gauge $a^\pm=0$.

In 3+1 QCD, we can not choose a gauge in which $a^\pm_i=0$. There are two
parts in the $a^\pm$-dependent part of the effective action. The first part
contains no $\epsilon$, and can be neatly included in \indep\ by replacing
$g^{-1}\partial_ig$ and $h^{-1}\partial_ih$ with $g^{-1}D_i^+g$ and
$h^{-1}D_i^+h$ respectively. Here we define $D^+_ig=\partial_ig+[a^+_i,g]$
and $D^-_ih=\partial_ih+[a^-_i,h]$.
The second part is proportional to $\epsilon$. Once again, we can drop out
some terms in this part and keep the most relevant one, which is
$${1\over g^2}\int d^2z 4\epsilon\hbox{tr}(a^+_iGa^-_iG^{-1}).$$
Finally, we add the $a^\pm$-dependent part to \indep\ and obtain the full
effective action
\eqn\full{S[g,h,G,a^\pm]={1\over g^2}\int d^2z\hbox{tr}[g^{-1}D^+_ig
Gh^{-1}D^-_ihG^{-1}-\epsilon(G^{-1}\partial_iG)^2+4\epsilon a^+_iGa^-_iG^{-1}]
.}

\newsec{$\theta$-$\phi$ correlator}

We show that reggeization of gluon fails to occur in the effective theory
described in the last section, in 2+1 QCD. It seems to us that calculations
in \cw\ concerning Reggeization are also valid
in 2+1 dimensions. Result of this section therefore poses the question whether
the ansatz \prop\ need be modified.  It remains an open problem whether
reggeization occurs in the effective theory in 3+1 QCD.

One
of the reasons for working with 2+1 QCD is that calculations are extremely
simplified, because the effective action is one dimensional and some terms
in \full\ are absent. Moreover, as we shall see in the Sect.5, the color
singlet part of \ampl\ can be calculated exactly and nonperturbatively with
the one dimensional effective action.

If there is only one transverse dimension, ``gauge fields'' $a^\pm$ can always
be gauge transformed into zero. The one dimensional effective action is
therefore
\eqn\oned{S[g,h,G]=\int dx\hbox{tr}[{1\over g^2}g^{-1}\dot{g}Gh^{-1}\dot{h}
G^{-1}-i{1\over e^2}(G^{-1}\dot{G})^2],}
where the dot denotes the derivative with respect to $x$, and $e^2=2g^2
\hbox{log}s/\pi$. It is important to notice that so far we have been working in
Minkowski spacetime, so there is a factor $i$ in the front of the action in
the path integral. The second term in \oned\ is pure imaginary. Together with
an overall $i$, it becomes negative in the exponential in the path integral,
as it should be. From now on we assume that the gauge group is $SU(N)$.
To do the perturbative calculation, we expand
\eqn\herm{g=e^{g\theta},\quad h=e^{g\phi},\quad G=e^{e\chi},}
The correlator of the two Wilson lines in the amplitude \ampl\ is expanded
\eqn\expan{\langle (g(x)-1)(h(0)-1)\rangle=g^2\langle \theta(x)\phi(0)\rangle
+{g^4\over 4}\langle\theta^2(x)\phi^2(0)\rangle+\dots.}

We are interested in reggeization of gluon which is weighted by a
factor $g^2$ and a function of $e^2$. So this is contained in the
first term representing the exchange of the color octet (for $SU(N)$
it is the exchange of adjoint multiplet). To check
reggeization, we have to calculate this term at least up to order $e^4$.
In other words, we have to calculate quantity $\langle\theta^a(q)\phi^b(-q)
\rangle$ up to two-loops, here we use the notation $\theta=\theta^aT_a$,
$\phi=\phi^aT_a$ and the anti-hermitian matrices are normalized by
$\tr T_aT_b=-\delta_{ab}$. For our purpose, it is enough to keep the
quadratic term in $\theta$-$\phi$ in the action \oned:
\eqn\quard{S=\int dx\hbox{tr}[\dot{\theta}G\dot{\phi}G^{-1}-{i\over e^2}
(G^{-1}\dot{G})^2].}

At the tree level, the correlator is
just the $\theta$-$\phi$ propagator $-(i/q^2)\delta_{ab}$ in \quard.
At the one-loop level, there are two Feynman diagrams as shown in Fig.1.
Solid lines in these diagrams are $\theta$-$\phi$ propagator, dotted lines
are $\chi$ propagator. Only two vertices out of infinite many vertices in
the action \quard\ appear in Fig.1.
These diagram are easily calculated. The
result is
\eqn\onlp{\langle \theta^a(q)\phi^b(-q)\rangle=-{i\over q^2}(1+{Ne^2\over 2}I)
\delta_{ab},}
up to one-loop. The integral $I$ is
\eqn\inte{I=\int{dk\over 2\pi}{1\over k^2+\mu^2},}
where we have introduced an infrared cut-off $\mu$. The reason for this
integral to arise is that all couplings are derivative couplings, as readily
seen from \quard. This integral is different
from the standard result \cw.
\eqn\stan{I_1={q^2\over 2}\int{dk\over 2\pi}{1\over (k^2+\mu^2)
((q-k)^2+\mu^2)}.}
Note that the usual calculation in 3+1 QCD is easily transformed into
2+1 QCD, so the integral \stan\ takes the same form as in \cw, except that
it is now an one dimensional integral. The difference between $I$ and $I_1$
is not crucial when $\mu$ is very small. Both of them can be calculated
exactly. $I$ differs from $I_1$ by $2\mu/(q^2+4\mu^2)$, which approaches
$\pi\delta(q)$ in the limit $\mu=0$. This difference is zero as long as
$q\ne 0$. So we conclude that up to one-loop, the effective action
\oned\ yields the same result as in the conventional calculation.

There are seventeen topologically distinct two-loop diagrams. We have drawn
fifteen of them in Fig.2. It is seen that at this order, all vertices up
to the six vertex as in 2.h are needed. The sum of diagrams
2a-2k is
$$-{5N^2e^4i\over 48 q^2}I^2\delta_{ab}.$$
The remaining diagrams in Fig.2 are obtained from those in Fig.1 by
replacing the $\chi$ propagator with its one-loop correction. There are two
additional such diagrams not drawn in Fig.2. These are similar to
2k and 2m except that two three vertices attached to the bottom line are
replaced by one four vertex. Since 2l cancels 2m, the other two
also cancel each other. The sum of 2n and 2o is
$$-{N^2e^4i\over 8q^2}(I^2+{I\over q^2}\int{dk\over 2\pi})\delta_{ab}.$$
The second integral in the parenthesis is ultraviolet divergent.
Conventionally this
integral is regularized to be zero. We shall see in sect.5 that
indeed in an exact calculation of the color singlet part of the correlation
of two Wilson lines, there is no
ultraviolet divergence. Dropping out this term and taking all results together,
the $\theta$-$\phi$ propagator up to two-loops is given by
\eqn\fail{\langle \theta^a(q)\phi^b(-q)\rangle=-{i\over q^2}(1-{Ne^2\over 2}
I+{11N^2e^4\over 48}I^2)\delta_{ab}.}
The reggeized gluon propagator is expected to be
$$-{i\over q^2}e^{-{Ne^2\over 2}I}\delta_{ab}$$
We thus see that result in \fail\ fails to give a reggeized gluon. In the
next section, we shall calculate $\theta$-$\phi$ with by a different method.
The result will be the same as in \fail. That method is very efficient and
we do not need to calculate so many diagrams as in Fig.2. Also, the method
will be used to do an exact calculation in sect.5. The infrared divergence
in $I$ is absent in a color singlet quantity.

\newsec{Another calculation of $\theta$-$\phi$ propagator}

It is technically desirable to develop a simple method to calculate quantities
such as $\theta$-$\phi$ correlator considered in the previous section, to
avoid calculating numerous diagrams. In this action we show that the
one dimensional action \quard\ can be easily transformed into a simpler action,
with which not only perturbative calculation is simplified, but exact result
is also available.

The idea is to introduce sources for $\theta$ and $\phi$
and integrate out these fields. To calculate $\theta$-$\phi$ correlator, we
start with action \quard. Introducing the source term
$$\int dx\hbox{tr}(\theta J_\phi+\phi J_\theta)$$
Now the correlator $\langle\theta^a(x)\phi^b(0)\rangle$ is written as
$$-{1\over Z}{\delta^2Z\over \delta J^a_\phi(x)\delta J^b_\theta(0)},$$
where the partition function with sources is obtained by integrating out
$\theta$ and $\phi$:
\eqn\parti{\eqalign{Z=&\int [dG]\hbox{Det}(G)\hbox{exp}[i\int dz\hbox{tr}
(-{i\over e^2}(G^{-1}\dot{G})^2\cr
&-\int dxdy \epsilon
(z-x)\epsilon(z-y)J_\phi(x)G(z)J_\theta(y)G^{-1}(z))],}}
where the determinant is defined by
$$\hbox{Det}(G)=\int[d\theta d\phi]\hbox{exp}(i\int dx\hbox{tr}(\dot{\theta}
G\dot{\phi}G^{-1}),$$
and the function $\epsilon(z)$ is the step function $\epsilon(z)=1/2$ when
$z>0$ and $\epsilon(z)=-1/2$ when $z<0$. An infrared cut-off factor
$\exp(-\mu |z|)$ is sometimes needed for this function.

It is easy to show diagrammatically that the above determinant gives rise to
an ultra-local term in the action, and therefore can be absorbed into the
definition of the measure $[dG]$. Hereafter we simply ignore this factor.
By a simple manipulation, we find that the correlator is given by
\eqn\usef{\langle\theta^a(q)\phi^b(-q)\rangle ={i\over q^2}\langle\hbox{tr}
(T_aGT_bG^{-1})(x)\rangle.}
The expectation value of $\hbox{tr}(T_aGT_bG^{-1})$ is independent of $x$ and
is defined with an action
\eqn\tdqcd{S(G)=-{i\over e^2}\int dx\hbox{tr}(G^{-1}\dot{G})^2.}
This action is the one dimensional analogue of the principle chiral model.
Interestingly, it can be derived from the two dimensional Euclidean QCD
on a cylinder, where $G$ is the holonomy around the spatial circle and $x$
is the Euclidean time. However, there is an important difference between
the 2D pure QCD and our model, i.e., in the 2D QCD a project operator
projecting out color non-singlet wave functions is inserted in the path
integral, while there is no such restriction in the model under consideration.
Nevertheless, our model can be solved exactly too. Let us calculate \usef\
perturbatively. Since we know that
$\langle\theta^a(q)\phi^b(-q)\rangle$ is proportional to $\delta_{ab}$, the
identity in \usef\ can be further simplified:
\eqn\sing{\langle\theta^a(q)\phi^b(-q)\rangle={i\over q^2}\delta_{ab}
{1\over N^2-1}
\langle\sum_a T_aGT_aG^{-1}\rangle=-{i\over q^2}\delta_{ab}
{1\over N^2-1}\langle\hbox{tr}G\hbox{tr}G^{-1}-1\rangle.}

Expanding $G$ in terms of $\chi$ up to order $e^4$, we have
\eqn\expand{{1\over N^2-1}\langle\hbox{tr}G\hbox{tr}G^{-1}-1\rangle
=1+{e^2\over N^2-1}\langle\hbox{tr}\chi^2\rangle+{e^4\over 4(N^2-1)}
\langle(\hbox{tr}\chi^2)^2\rangle+{Ne^4\over 12(N^2-1)}\langle\hbox{tr}
\chi^4\rangle.}
Expectation values on the r.h.s. of the above equation can be easily
calculated. We calculate the second term as an example.
\eqn\plug{{Ne^2\over N^2-1}{1\over\delta(0)}\int{dk\over 2\pi}\langle
\chi^a(k)\chi^a(-k)\rangle,}
where the delta function in the denominator comes from the observation
that the expectation value $\langle\tr\chi^2(x)\rangle$ is independent
of its position, and $\delta(0)$ is just the cut-off of the volume. To the
first order, the correlator on the l.h.s. of the above equation is just
$$\delta(0)\sum_a\delta_{aa}{1\over 2(k^2+\mu^2)}=\delta(0)(N^2-1){1\over
2(k^2+\mu^2)}$$
Plugging this result back into \plug, we obtain the result $-(Ne^2/2) I$, in
agreement with result obtained in the last section. So this simple result
obtained from the $\chi$ propagator summarizes two diagrams in Fig.1. Since
we are interested in the correlator in \sing\ up to the order $e^4$, we have
to calculate the next order contribution to \plug. This is given by the
diagram in Fig.3. This single diagram summarizes diagrams 2n and 2o
in Fig.2. Plug this quantity into \plug,
\eqn\plugg{{Ne^2\over N^2-1}\langle \hbox{tr}\chi^2\rangle=-{Ne^2\over 2}I
+{N^2e^4\over 8}I^2.}
We find that, had there been not for the last two terms in \expand, this
formula would give us reggeization. The last two terms can be
calculated similarly. Up to the order $e^4$, they are
$$\eqalign{&{e^4\over 4(N^2-1)}\langle(\hbox{tr}\chi^2)^2\rangle
={(N^2+1)e^4\over 16}I^2,\cr
&{Ne^4\over 12(N^2-1)}\langle\hbox{tr}\chi^4\rangle={(2N^2-3)e^4\over 48}I^2.}
$$
These two terms together summarize diagrams 2a-2k. Taking results obtained
into \expand
\eqn\pert{{1\over N^2-1}\langle\hbox{tr}G\hbox{tr}G^{-1}-1\rangle
=1-{Ne^2\over 2}I+{11N^2e^4\over 48}I^2+\dots.}
This calculation is in full agreement with \fail, which was obtained by
tediously counting many diagrams.

\newsec{The color singlet correlator: exact result}

Although we have not seen reggeization of gluon, this does not imply that
Verlindes' approach is wrong. What we really have to calculate is color
singlet amplitude, and this amplitude will give rise to the hadron scattering
amplitude when properly folded by hadron wave functions \ref\levin{
for example, see L.N. Lipatov, in Perturbative Quantum Chromodynamics,
ed. A.H. Mueller, World Scientific, 1989; and E.M. Levin, Orsay lectures
on low x deep inelastic scattering, 1991.}. From \expan, the proper quantity
to consider is
\eqn\corre{{g^4\over 4}\langle\hbox{tr}\theta^2(x)\tr \phi^2(0)\rangle.}
This is the leading term in the expansion \expan\ when we consider only
color singlet combination. High order terms in \expan\ should be included
in principle, but it is inconsistent to do this with action \quard\ or even
the unsimplified action \act, because higher order terms in $g^2$ in the action
are omitted. However, \corre\ contains all orders in $e^2$.

We shall calculate color singlet correlator \corre\ exactly with the action
\quard\ in this section, this will constitute our main result in this paper.
To do the exact calculation, we introduce sources as in \parti, then the
correlator \corre\ is written as
$$\langle\tr\theta^2(x)\tr\phi^2(0)\rangle={1\over Z}{\delta^4Z\over
\delta J^a_\phi(x)\delta J^a_\phi(x)\delta J^b_\theta(0)\delta J^b_\theta(0)
},$$
and this formula leads to, after integration over $\theta$ and $\phi$
$$\langle\tr\theta^2(x)\tr\phi^2(0)\rangle=-2\int dzdz'\epsilon(z-x)
\epsilon(z'-x)\epsilon(z)\epsilon(z')\langle G^{ab}(z)G^{ab}(z')\rangle,$$
where $G^{ab}=\tr (T_aGT_bG^{-1})$. Using some identities for unitary
matrices, the above expression can be further simplified:
\eqn\calc{\eqalign{\langle\tr\theta^2(x)\tr\phi^2(0)\rangle=&-2\int dzdz'
\epsilon(z-x)\epsilon(z'-x)\epsilon(z)\epsilon(z')\cr
&\langle\tr\left( G^{-1}(z')G(z)\right)
\tr\left( G^{-1}(z)G(z')\right)-1\rangle.}}
Again, the correlator on the r.h.s. of the above equation should be defined
with action \tdqcd. Before we set off to calculate this correlator, we
remark that calculation of a general correlator, such as the one discussed
in the last section, is very involved, although the particular correlator
in \calc\ is surprisingly easy to calculate.

By translational invariance, we set $z'=0$. We also assume $z>0$, because
the correlator is an even function of $z$. We discretize the $z$ line, the
Euclidean action (after a factor $i$ in the path integral is absorbed
into the action) is
\eqn\discr{S(G)=-{1\over e^2\Delta z}\sum_n\tr(G^{-1}_nG_{n+1}-1)^2,}
where $\Delta z$ is the spacing between two adjacent sites. Define new
variable $U_n=G^{-1}_nG_{n+1}$, then $\tr \left(G^{-1}(0)G(z)\right)=
\tr (U_1\dots U_L)$ and $\tr\left( G^{-1}(z)G(0)\right)=\tr (U^+_L\dots U^+_1)
$. Action \discr\ becomes ultra-local in terms of new variables $U_n$, this
is why the correlator is easy to calculate. It is easy to see that we need
only to calculate
\eqn\chief{\int U_{ij}U^+_{lk}\exp({1\over e^2\Delta z}\tr(U-1)^2)dU.}
Once this quantity is known, we replace $U$ by $U_N$ and insert it into the
correlator we want to calculate and obtain a recursion relation. Since
spacing $\Delta z$ can be arbitrarily small, it is sufficient to calculate
quantity \chief\ up to the first order in $e^2\Delta z$. This can be
readily done and we only give the final result:
\eqn\final{\int U_{ij}U^+_{lk}\exp({1\over e^2\Delta z}\tr (U-1)^2)dU=
(1-{Ne^2\over 2}\Delta z)\delta_{ij}\delta_{lk}+{e^2\over 2}\Delta z
\delta_{ik}\delta_{jl},}
where we assume that the integral $\int\exp({1\over e^2\Delta z}\tr(U-1)^2)dU$
is normalized to one. Defining
$$F(z)=\langle\tr(U_1\dots U_L)\tr(U^+_L\dots U^+_1)\rangle,$$
and inserting into it \final\ with $U=U_L$, we obtain $F(z)=(1-{Ne^2\over 2}
\Delta z)F(z-\Delta z)+{Ne^2\over 2}\Delta z$ or the differential equation
$${dF(z)\over dz}=-{Ne^2\over 2}F(z)+{Ne^2\over 2}.$$
The general solution to this equation is $F(z)=1+C\exp(-{Ne^2\over 2}z)$,
$z>0$. To fix the coefficient, let $z=0$, from the definition of $F(z)$
we know that $F(0)=N^2$. Thus $C=N^2-1$ and $F(z)=1+(N^2-1)\exp(-{Ne^2\over
2}|z|)$, here $z$ can be either positive or negative. Substituting this
exact result into \calc, we obtain our main result,
\eqn\main{\langle\tr\theta^2(x)\tr\phi^2(0)\rangle=-2(N^2-1)\int dzdz'
\epsilon(z-x)\epsilon(z'-x)\epsilon(z)\epsilon(z')\exp(-{Ne^2\over 2}|z-z'|).}

The rest of this section is devoted to a discussion and comparison of the
above exact result to perturbative consideration. Physical implications of
this result will be discussed in the next section.

First transforming \main\ into momentum space, using $\epsilon(k)=\int dx
\exp(-ixk)\epsilon(x)=-ik/(k^2+\mu^2)$, $\mu$ is a infrared cut-off parameter,
we obtain
\eqn\momen{\eqalign{&\langle\tr\theta^2(q)\tr\phi^2(-q)\rangle=
-2(N^2-1)\int {dk_1dk_2\over (2\pi)^2}\epsilon(k_1)\epsilon(q-k_1)\epsilon(k_2)
\epsilon(q-k_2)\cr
&\int dz\exp(i(k_1-k_2)z -{Ne^2\over2}|z|)\cr
&=-2(N^2-1)\int{dk_1dk_2\over (2\pi)^2}\epsilon(k_1)\epsilon(q-k_1)\epsilon
(k_2)\epsilon(q-k_2){Ne^2\over (k_1-k_2)^2+N^2e^4/4}.}}

We now see how the above exact result reproduces perturbative result in the
first two orders. Let $e^2=0$ in the first equality in \momen, integration
over $z$ yields a delta function $\delta(k_1-k_2)$. So to the first order,
the correlator is
$$\langle\tr\theta^2(q)\tr\phi^2(-q)\rangle=-2(N^2-1)\int {dk\over 2\pi}
{1\over (k^2+\mu^2)((q-k)^2+\mu^2)}.$$
This result is the same as given by diagram in Fig.4a. The factor $N^2-1$
come from taking trace, and the factor $2$ is a symmetric factor. The above
integral is infrared divergent, although the full result in \momen\ is
finite. Take the next order, and do the integral
$$\int dz|z|e^{i(k_1-k_2)z}={2(\mu^2-(k_1-k_2)^2)\over ((k_1-k_2)^2+\mu^2)^2},
$$
where we introduced a cut-off factor in the integration. Because of the
infrared singularity in the above formula, we can not simply drop out the
$\mu^2$ term in the
numerator. Plugging the above formula into the first equality in \momen,
it is easy to see, after some simple calculation, that the next order
contribution to the correlator is exactly the same as coming from the
H-diagram in Fig.4b and two loop corrections to Fig.4a. The H-diagram
summarizes many contributions in conventional calculation, and the result
is written in terms of Lipatov's emission vertex \lip.
We thus verified that our exact result agrees with perturbative calculation.

The integral in the second equality of \momen\ is like a generalization of
the H-diagram in Fig.4a, with an effective massive $G$-propagator. The mass
is $m=Ne^2/2$. Note that in 2+1 dimensions the coupling constant $g^2$,
therefore $e^2$ indeed has a mass dimension. Here physics is drastically
different from that in 3+1 dimensions, where there could be no such mass
arising, since the coupling constant there is dimensionless. Finally, we
perform the integral in \momen\ and obtain
\eqn\close{\langle\tr\theta^2(q)\tr\phi^2(-q)\rangle=-{2(N^2-1)\over m(q^2+
m^2)}+(N^2-1)\left(-{\pi^2\over m}\delta^2(q)+{2\pi\over m^2}\delta(q)
\right).}
The first term in the above formula tells us that the color singlet
correlator is effectively due
to exchange of a single massive ``pomeron''. Indeed, combining this result with
a factor $g^4/4$ as in formula \corre, we find that the coupling constant
between quark and the pomeron is $g^2/\sqrt{2m}=g\sqrt{{\pi\over 2N\log s}}.$
Since this coupling constant depends on the total energy, the ``pomeron''
has no simple 2+1 spacetime interpretation. The second term in \close\
unvanishes only when $q=0$.

\newsec{Discussion}

\subsec{Hadron-hadron scattering in 2+1 QCD}

Upon using wave functions of hadrons, and quark-quark and quark-anti-quark
scattering amplitudes, amplitude and cross section of hadron-hadron
scattering can be calculated. We calculated in the last section the
color singlet part of quark-quark scattering amplitude. It remains to be
done the quark-anti-quark scattering amplitude. We now show that as for
its color singlet part is concerned, the result is the same as for
quark-quark scattering amplitude. According to \nach, quark-anti-quark
scattering amplitude reduces at high energy to a similar formula as in \ampl,
except that $V_-$, the Wilson line for the right-moving quark, is replaced
by $V_-^*$ for the right-moving anti-quark. This Wilson line is just
$h^*$. So the color singlet part of quark-anti-quark scattering amplitude
is given by
\eqn\antiq{A(s,t)={is\over 2m^2_q}{g^4\over 4}\int dze^{-iqz}\langle\
\tr\theta^2(z)\tr(\phi^*)^2\rangle.}
Since $\phi$ is an anti-hermitian matrix, $\tr(\phi^*)^2=\tr(\phi^t)^2$,
$\phi^t$ is the transpose of $\phi$. By definition of trace, $\tr(\phi^t)^2
=\tr\phi^2$. Thus, the quark-anti-quark scattering amplitude is identical
to quark-quark scattering amplitude, if we are interested only in color
singlet.

Consider now hadron-hadron scattering amplitude. Since hadrons are color
singlet states, physical processes only involve exchange of color singlet
objects. The simplest possible exchange is two gluons, and there are also
many corrections to this process. What we calculated in the last section
is just these corrections up to all orders in $e^2$. Let us calculate
elastic scattering amplitude of hadron $A$ and hadron $B$. Assume
that the effective vertex for hadron $A$ to emit two virtual gluons is
$\Phi_A^{\mu_1\mu_2}(k_1^2, (q-k_1)^2,s_1)\delta_{ab}$, where $s_1=-2k_1\cdot
p_A$; and the effective vertex for
hadron $B$ is $\Phi_B^{\nu_1\nu_2}(k_2^2,(q-k_2)^2,s_2)\delta_{cd}$, where
$s_2=2k_2\cdot p_B$. The scattering
amplitude is
\eqn\has{\eqalign{A(s,t)&=\int{d^3k_1d^3k_2\over (2\pi)^6}\Phi_A^{\mu_1\mu_2}
(k_1^2,(q-k_1)^2,s_1)
\Phi_B^{\nu_1\nu_2}(k_2^2,(q-k_2)^2,s_2)(2{p_B^{\mu_1}p_A^{\nu_1}\over s})
(2{p_B^{\mu_2}
p_A^{\nu_2}\over s})\cr
&A_{aa,bb}(k_1,q-k_1,k_2,q-k_2),}}
where $A_{ab,cd}$ is the amplitude generalizing two gluon propagators.
Following \levin, we perform the integral over longitudinal
momenta, the resulting integrand depends only on transverse momenta. We
define the following ``structure functions'' depending only on transverse
momenta:
\eqn\struc{\eqalign{\phi_A(k,q)&=\int_L{ds_1\over 2\pi i}{p_B^\mu p_B^\nu
\over s^2}\Phi_A^{\mu\nu}(k^2,(q-k)^2,s_1),\cr
\phi_B(k,q)&=\int_L{ds_2\over 2\pi i}{p_A^\mu p_A^\nu\over s^2}\Phi_B^{\mu
\nu}(k^2,(q-k)^2,s_2).}}
Contours are chosen to avoid the right and left cuts on the real axis. For
details please consult \levin. The hadron-hadron amplitude is
\eqn\redu{A(s,t)=2is\int{dk_1dk_2\over (2\pi)^2}\phi_A(k_1,q)\phi_B(k_2,q)
A'_{aa,bb}(k_1,k_2,q),}
where $A'_{aa,bb}$ can be read off from the color singlet part of quark-quark
amplitude. Applying the same procedure to quark-quark scattering, we would
obtain a similar formula as \redu, now the effective vertex for quark to
emit two gluons is proportional to $(T_aT_b)_{AB}$ (consider color singlet
only). Taking trace over $A,B$,
it is just $-\delta_{ab}$. Thus, from \momen, we find
\eqn\sandw{A'_{aa,bb}(k_1,k_2,q)=(N^2-1)\epsilon(k_1)\epsilon(q-k_1)
\epsilon(k_2)\epsilon(q-k_2){Ne^2\over (k_1-k_2)^2+N^2e^4/4}.}
This amplitude is real. We have taken another sign in \ampl\ into account.
Note that in \has\ only two gluon exchange and its
corrections are taken into account, while \sandw\ contains more terms,
such as exchange of three gluons, see \cw. Nevertheless \redu\ still works
in this general case, provided we properly modify $\phi_A$ and $\phi_B$.

The integral in \redu\ has no divergence at $k_1=0$
or $k_2=0$, for $\phi_A$ or $\phi_B$ is equal to the difference of values of
a function taking at $k_1$ and $k_1=0$. $\phi_A$ is a symmetric function
of $k_1$ and $q-k_1$, so there is no divergence at $k_1=q$. There is no
divergence in \redu\ at $k_1=k_2$ either, since summation of all orders
cures the infrared problem. When the energy is extremely high, function
$${Ne^2\over (k_1-k_2)^2+N^2e^4/4}$$
does not depend on $k_1-k_2$ sensitively for a large range of momenta, so
it can be effectively replaced by $4/(Ne^2)$. The amplitude \redu\ is
approximated by
\eqn\appro{A(s,t)={8is(N^2-1)\over Ne^2}\int{dk_1dk_2\over (2\pi)^2}
\phi_A(k_1,q)\phi_B(k_2,q)
\epsilon(k_1)\epsilon(q-k_1)\epsilon(k_2)\epsilon(q-k_2).}
The double integral effectively factorizes into two independent integrals.
The double integral is dimensionless, since each structure function $\phi$
contains a factor $g^2$ therefore has a mass dimension. $s$ is measured
by a certain mass, thus it is dimensionless. We conclude that the amplitude
$A(s,t)$ has a dimension of $1/e^2$, the length dimension, as it should be
in 2+1 dimensions. Apart from a factor $s$, the amplitude decreases as
$1/e^2\sim 1/\log s$ at high energies. This physics is quite different
from what happens in 3+1 dimensions, where one expects that apart from $s$,
there
is an additional Regge-like factor $s^\alpha$, $\alpha>0$. The absence of
this factor in 2+1 dimensions is reasonable: there is no way to construct
a dimensionless parameter $\alpha$ out a dimensionful parameter $g^2$.

The amplitude \redu\ can be used to calculate
the elastic cross section $\sigma_e$. In 2+1 dimensions, the formula of
cross section at high energy is given by
$$\sigma_e={1\over 8\pi s^2}\int dq|A(s,t)|^2,$$
where the integral is over the transverse component of the momentum transfer.
{}From the above formula and \appro, we find that the elastic cross section
falls off like $1/e^4\sim 1/(\log s)^2$. The total cross section, by the
optical theorem, is given by
$$\sigma_t\sim {1\over s}\hbox{Im} A(s,t=0).$$
Thus, the total cross section falls off like $1/e^2\sim 1/\log s$.

In conclusion, we have shown that the high energy behavior of hadron-hadron
scattering in 2+1 differs drastically from that in 3+1 dimensions. This is
not surprising for a couple of reasons. First, the coupling constant in 2+1
QCD is dimensionful, so the theory is super-renormalizable. This property is
reflected in the one dimensional effective action for high energy
scattering, there is simply no ultraviolet divergence. For the same reason,
there is an effective mass gap $Ne^2$ arising in high energy scattering.
Second, there is only one transverse dimension in 2+1 spacetime, so there
is no arbitrarily high transverse angular momentum. As a result, one does
not expect Regge behavior arises, which is due to exchange of states of
high spins. The lesson from the study of this paper is that one should be
extremely careful in extending any results obtained in lower dimensions
to full four dimensions.

\subsec{Some comments on 3+1 QCD}

We have used the trick of integrating out the $\theta$ and $\phi$ fields
to solve our problem in 2+1 QCD. This very same trick can be generalized
to 3+1 QCD, starting with the effective action \full. Up to the leading
order in $g^2$, which is what can be done consistently in Verlindes'
approach, action \full\ becomes
\eqn\comm{S[\theta,\phi,G,a^\pm]=\int d^2z\tr[D^+_i\theta GD^-_i\phi
G^{-1}-{i\over e^2}(G^{-1}\partial_iG)^2+{i\over e^2}a^+_iGa^-_iG^{-1}],}
where we expanded $g$ and $h$ as in \herm. To integrate out $\theta$
and $\phi$, we do transformation
$$\phi\rightarrow G^{-1}\phi G,\quad a^-_i\rightarrow G^{-1}a^-_iG
+G^{-1}\partial_iG .$$
Now the first term in \comm\ becomes $\tr(D_i^+\theta D_i^-\phi)$. Integrating
out $\theta$ and $\phi$, we obtain the effective action for $G$ and $a^\pm$
\eqn\effect{S[G,a^\pm]=i\tr\log[-D_i^+D_i^-]-{i\over e^2}\int d^2z\tr
\left((G^{-1}\partial_iG)^2-a_i^+(a_i^-+\partial_i G G^{-1})\right),}
where the determinant is defined with respect to adjoint fields $\theta$
and $\phi$.
Now the Wilson line correlation function \expan\ is to be calculated as
a correlator involving $G$ and $a^\pm$ with the above effective action.

So far we have not chosen a gauge. On might choose the Landau gauge
$\partial_ia^\pm_i=0$, so $a^\pm_i=\epsilon_{ij}\partial_j\alpha^\pm$.
The effective action \effect\ for $G$ and $\alpha^\pm$ can be evaluated.
The first term in \effect, the term resulting from integrating out
$\theta$ and $\phi$, will be similar to a Wess-Zumino-Witten model.
This term, we believe, is responsible for the major difference in
physics between 3+1 QCD and 2+1 QCD. When the total energy is extremely
high, the second term in \effect\ is much smaller compared to the first term,
since $e^2$ is large as it contains a factor $\log s$. The first term has no
scale and therefore is conformally invariant. We thus expect that the leading
contribution to the hadron-hadron scattering contains no mass gap.
This agrees qualitatively with conventional leading log approach. However,
one can not ignore the second term in \effect, for it contains the kinetic
term for $G$. Therefore, $e^2$ thus $\log s$ must enter the calculation
of scattering amplitude.
We leave a detailed study of the model defined by \effect\ to the future.

\noindent{\bf Acknowledgments} We thank J.R. Cudell, E.M. Levin and H.
Verlinde for interesting discussions. This work was supported by DOE
contract DE-FG02-91ER40688-Task A.

\listrefs
\vfill\eject
\centerline{\bf Figure captions}
\vskip5mm

\noindent Fig.1 One-loop diagrams contributing to $\theta$-$\phi$ correlator.

\noindent Fig.2 Two-loop diagrams contributing to $\theta$-$\phi$ correlator.

\noindent Fig.3 One-loop correction to the $\chi$ propagator.

\noindent Fig.4 The box diagram fig.4a and the H-diagram fig.4b, contributing
to the
color singlet part of quark-quark scattering at the first order and the second
order, respectively.
\end